\documentclass{aa}

\usepackage{color}

\usepackage{graphics}
\usepackage{amssymb,amsmath}
\usepackage{amsmath}
\usepackage{psfrag}
\usepackage{graphicx}

\usepackage{float}
\usepackage[caption = false]{subfig}

\usepackage{graphics}
\usepackage{amssymb,amsmath}
\usepackage{amsmath}
\usepackage{psfrag}
\usepackage{graphicx}
\usepackage{natbib}

\newcommand{\ve}[1]{\mathbf{q1}}

\newcommand{\be}{\begin{equation}}      
\newcommand{\ee}{\end{equation}}      
      
\newcommand{\bef}{\begin{figure}}      
\newcommand{\eef}{\end{figure}}      
\newcommand{\bea}{\begin{eqnarray}}    
\newcommand{\eea}{\end{eqnarray}}

\begin{document}

\title{Properties of self-gravitating quasi-stationary states} 
  
\titlerunning{Properties of self-gravitating quasi stationary states}
  
\authorrunning{Sylos Labini and Capuzzo-Dolcetta }  
  
  \author  {Francesco   Sylos  Labini  \inst{1,2,3} and 
  Roberto Capuzzo-Dolcetta   \inst{4,1}  
   }

        \institute{
         CREF, Centro Ricerche Enrico Fermi,  
         Via Panisperna 89A, I-00184, Roma, Italy  
             \and 
             Istituto dei Sistemi Complessi, Consiglio Nazionale delle Ricerche, I-00185
          Roma,  Italy  
          \and   
          Istituto  Nazionale  Fisica  Nucleare,
          Dipartimento  di  Fisica, Universit\`a  ``Sapienza'',  I-00185
          Roma,  Italy
           \and  Dipartimento di Fisica, Sapienza, Universit\'a di
  Roma, piazzale Aldo Moro 2, I-00185, Roma, Italy     \\}

\date{Received / Accepted}

\abstract{
Initially far out-of-equilibrium,  self-gravitating systems form   quasi-stationary states (QSS) through
a  
collisionless relaxation dynamics. 
These may arise from a bottom-up aggregation of structures or in a top-down frame;
their quasi-equilibrium properties are well described by the Jeans equation and 
{are not universal. These QSS depend on initial conditions. }
To understand the origin of such dependence, 
we present the results of numerical experiments  of  initially cold and spherical systems characterized by 
various choices of the spectrum of initial density fluctuations. The amplitude of {such} fluctuations  determines whether the system relaxes in a top-down or bottom-up manner.
We find that  statistical properties of the resulting QSS  mainly 
depend upon the amount of energy exchanged  during the formation process. 
In particular, in the violent top-down collapses  the energy exchange is large and the QSS show 
an inner core with an almost flat density profile  and a quasi Maxwell-Boltzmann (isotropic) 
velocity distribution,
while their  outer regions display a density profile $\rho(r) \propto r^{-\alpha}$ ($\alpha >0$) with radially 
elongated orbits. We show analytically that $\alpha=4$,
in agreement with numerical experiments. 
In the less violent 
bottom-up dynamics,
the  energy exchange is much smaller,
the orbits are less elongated, and $0< \alpha(r) \le 4$, where the 
density profile is well fitted by the Navarro-Frenk-White behavior. 
Such a dynamical evolution
is shown by both  nonuniform spherical isolated systems and  by
  halos extracted from cosmological simulations.
We consider the relation of these results with the core-cusp problem 
and conclude that this can be
solved naturally if galaxies form through a monolithic collapse. 
}
  \maketitle


\section{Introduction} 
\label{sect:intro}

Unlike typical short-range interacting systems, which tend to relax to 
thermodynamical equilibria through collisions,  
long-range interacting systems
are driven to quasi-equilibrium configurations, or  quasi-stationary states (QSS), 
by a mean-field collisionless relaxation dynamics. These QSS are close to virial 
equilibrium and their lifetime diverges with the number of particles $N$ because of the decreasing effect of two-body encounters  
\citep{lyndenbell_1967,Padmanabhan:1989gm,Dauxois_etal_2002,Campa_etal_2014,Levin_etal_2014,RCDbook}.
Comprehension of the statistical properties of the QSS requires the understanding of how their dynamics determines the evolution of the system from an out-of-equilibrium condition to a quasi-stationary configuration.

If the long-range force at work is self-gravity, QSS are reached in two different frames, corresponding to
that of a finite and isolated system or to that of a system ``embedded'' in an expanding space, respectively, this latter representing a typical cosmological case. In both cases, 
depending on the  initial conditions (IC), 
the dynamics may correspond to a {top-down}
monolithic collapse or to a {bottom-up} aggregation of substructures. In this work we single out the properties of the IC by 
 determining which evolutionary path brings 
 a certain configuration towards a QSS, regardless of whether the system is finite and isolated or embedded in an infinite expanding background.
 

{The top-down monolithic collapse is modeled, in the simplest way, as the progression of a gravitational instability out of the linearity because such a collapse happens when the over-density is such that the self-gravity dominates over the expanding background  \citep{Peebles_1983}. If the amplitude
of a local over-density is
large enough  then
 tidal effects of neighboring density perturbations can be neglected and its evolution proceeds, essentially, as that of an isolated  perturbation. These are precisely the hypotheses 
 assumed in 
the nonlinear, gravitational instability model, known as the spherical collapse model, which is analytically solvable and paradigmatic \citep{sahni_95}.
 Indeed, the evolution of an isolated over-density in an expanding background should
reproduce, in physical coordinates, that obtained in open boundary conditions without expansion
\citep{joyce+syloslabini_2013}.}
The collapse and stabilization of such an over-density  has been 
studied since the first numerical experiments with self-gravitating systems in both isolated and embedded cases 
\citep{henon_1964,vanalbada_1982,aarseth_etal_1988,aguilar+merritt_1990,theis+spurzem_1999,   boily_etal_2002, roy+perez_2004, boily+athanassoula_2006, barnes_etal_2009,jmsl_2009, syloslabini_2012, worrakitpoonpon_2014, merritt+aguilar_1985, aguilar+merritt_1990, theis+spurzem_1999,syloslabini_2013, SylosLabini+Benhaiem+Joyce_2015, Benhaiem+SylosLabini_2015,Benhaiem_etal_2016, Benhaiem+SylosLabini_2017,SpeRCD2017}.

On the other hand, a QSS can be originated from a 
bottom-up hierarchical aggregation process,
in which smaller substructures merge to form larger and larger systems. 
If the system is infinite this process continues without ending, while if the system is finite the aggregation eventually halts. 
Bottom-up structure formation is typical of 
standard cosmological scenarios, like the cold dark matter (CDM) scenario because of the long-range nature of density correlations \citep{Blumenthal_etal_1982,Bond_etal_1982,Peebles_1983,Blumenthal_etal_1984}.
{In the case of cosmological systems, 
if the velocity dispersion is large then the collapse occurs 
for  objects that are big enough to make their gravitational potential
overcoming the pressure  due to random motions, which corresponds to the so-called
hot dark matter scenarios. 
In this latter case, density correlations have a sharp cutoff
beyond a scale corresponding to the size
of the perturbations that first 
become nonlinear \citep{Peebles_1983}.}

The statistical properties of the QSS depend on which of the 
two evolutionary paths described above was followed by the system in exam. 
In particular, in this paper we show that these properties are essentially related to the 
violence, in terms of the particle energy variation, of the process leading toward settling the system in a QSS. This process is very quick in the case of a top-down monolithic collapse, while slower for a hierarchical, bottom-up, aggregation process.

The focus of our study is the investigation of QSS with power-law density profiles and for this reason we consider cold IC that correspond to 
 far out-of-equilibrium configurations
(i.e., with a virial ratio\footnote{The virial ratio $q$ is defined in this work as $Q\equiv \frac{2K}{W}$, where $K$ and $W$ are the total kinetic and potential energy, respectively.} $|Q| \ll 1$:
{if $|Q| \approx 1$) then the collapse is inhibited and the system relaxes gently to form a compact core 
with an exponentially decaying density profile \citep{syloslabini_2013,Benetti_2014}.}

In order to study the two dynamical mechanisms outlined above, in this paper we consider, through numerical $N$-body experiments, the evolution of simple systems  
corresponding to finite spherical distributions with
various initial density fluctuations power spectra.
Changing the amplitude of such fluctuations 
allows us to pass from a top-down to a bottom-up process, and thus to explore 
the full  dynamics phase-space. This study  aims to develop a unified understanding of the properties of the QSS generated by both dynamical mechanisms. 
We also consider the properties of QSS formed in cosmological N-body simulations, that is, the so-called halos. To this purpose, we consider halos extracted from the Abacus simulations 
\citep{Garrison_2018,Garrison_2019}, where a 
CDM scenario is adopted. We show that their properties can be understood in the same theoretical framework developed above
{and discuss the reason for such a case}. 

The paper is organized as follows. 
In  in Sect.\ref{sect:simulations} we present the main characteristics of our 
$N$-body experiments  for  isolated systems we considered and of those of the Abacus simulations leading to cosmological halos.
In Sect.\ref{sect:scm+halos} we discuss the case of  models of uniform and 
nonuniform spherical collapses
that show the transition from a bottom-up to a top-down clustering.
The properties of cosmological halos are also considered.
The astrophysical implications of our findings are discussed in Sect.\ref{sect:concl}, in which 
we also draw our conclusions.


\section{Models and methods}
\label{sect:simulations}

\subsection{Isolated systems}

We considered two types of IC in our numerical experiments of finite systems. 
The first is represented by  spherical, isolated, spatially homogeneous,  and cold over-densities   
of $N$ particles 
  of mass $m$ with zero initial velocity in which particles are 
randomly distributed, that is, have Poisson density  fluctuations,
\[
\delta = \sqrt{ \frac{\langle \Delta N^2 \rangle}{\langle N \rangle^2}}     \sim N^{-1/2} \;. 
\]
In order to explore the role of fluctuations 
we let $N$ vary in the range $10^4-10^6$ while the total mass
and size of the system are taken as constant. We chose a 
normalization to an astrophysical object; the total mass is  $M=10^{10} M_\odot$ and the initial
radius is  $R_0=100$ kpc, therefore the free-fall time 
\be
\label{tauff}
\tau_{ff} \approx \frac{1}{\sqrt{G\rho}} \approx 1.5  \;\; \mbox{Gyr,}
\ee
where $\rho = 3M/(4 \pi R_0^3)$ is the system density.
{As mentioned in the introduction, the most violent evolution occurs, of course,  when the initial velocity 
dispersion is zero ($Q=0$); when the initial virial ratio is in the range 
$-0.5<Q<0$ then the collapse is less violent but, qualitatively, the dynamical evolution remains 
the similar to that of  $Q=0$ \citep{syloslabini_2013}. For warmer IC (i.e., $-1<Q<-0.5)$ the collapse is halted by the effect of the large velocity dispersion and the system reaches a configuration characterized by a compact core a very diluted halo \citep{Benetti_2014}.
We focussed our attention 
on  cold IC because a nontrivial power-law density profile is found to develop for those cases alone.}

The second family of IC is still represented by isolated, almost spherical, spatially  homogeneous  and cold systems that have the same $M,R_0$ as before, but that have initial density fluctuations that are larger than Poisson fluctuations.
 These are generated 
by randomly distributing $N_c$ points in a sphere of radius $R$. Each of these 
points is then considered as a center of a spherical subsystem of 
$N_p$ particles that are also randomly distributed in a smaller spherical volume.
 We take  the radius of each subsystem to be 
$r_s= 2 \Lambda_c$, where $\Lambda_c$ is the average distance between the $N_c$ 
particles(i.e., $\Lambda_c = 0.55 (3N_c/3\pi R_0^3)^{1/3}$). 
In this way a moderate overlap between different subclumps is allowed
to smooth out initial fluctuations when $N_c$ is sufficiently large(i.e., $N_c>10^2$).
The total number of particles is thus $N=N_c \times N_p \approx10^6$. 
A realization can be  characterized by the parameter 
\be
\label{gamma}
\gamma = \frac{N}{N_c} \,
,\ee
where we chose $\gamma \in [10,10^5],$ where 
$\gamma= 10^5$ for the initially strongly clustered case
and $\gamma=10$ for the less clustered case. 
Indeed, the smaller  $N_c$ is, the larger $\gamma$ is and the 
larger the initial fluctuations $\delta \sim N_c^{-1/2} \sim (\gamma/N)^{1/2}$
in sufficiently large scale $r > \Lambda_c$.
The initial velocities are  taken to be zero as in the previous case.

 \subsection{The code}

We performed all our simulations by means of the publicly available and widely used
code {\tt Gadget-2} \citep{Springel_2005}. 
The gravitational interaction is evaluated by direct summation over close neighbors 
and via a multipolar expansion on a larger scale. 
In this way, the number of computations is sensibly lower compared to the usual 
$N^2$ scaling, which is characteristic of the direct-summation $N$-body algorithms.
The gravitational interaction on the small distance scale is
regularized with the so-called gravitational softening
$\varepsilon$. The force has its purely Newtonian value 
at separations greater than $\varepsilon$ ($r\geq \varepsilon$) 
while it is smoothed at shorter separations. 
The assumed functional form of the regularized potential 
is a cubic spline interpolating between the exact Newtonian 
potential at $r = \varepsilon$ and a constant value at $r=0,$ 
where the mutual gravitational force vanishes (the exact expression can be found in \cite{Springel_2005}). 
A detailed study of the parameter space of the code {\tt Gadget-2},  
for simulations considering only Newtonian gravity, has been reported in
\cite{jmsl_2009,syloslabini_2013}.
In the simulations that we discuss in what follows we always 
keep energy, momentum, and angular momentum conservation at a level of precision better than $1 \%$. 

The criterion for our choice of softening length $\varepsilon$  is 
that this is sufficiently small so
the numerical results are independent of it, and we interpret
our results as being representative of the limit $\varepsilon=0$.
A convergence study by varying  $\varepsilon$ is presented in \cite{jmsl_2009}, where 
it was concluded that 
 results are  $\varepsilon$ independent as long the minimal 
 radius of the system  $R_{min} $ during the collapse is 
larger than  $\varepsilon$.  We take $\varepsilon=0.05$ kpc
but we also considered experiments with $\varepsilon=0.005$ kpc.
Given that collisional effects are negligible, occurring on much longer timescales 
than the collapse characteristic timescale,  
this result can be understood as due to the system mean-field,
whose variation is the source of the dynamics, 
remains Newtonian as long as $R_{min} \gg  \varepsilon$.

\subsection{Cosmological halos}

We analyzed several  halo catalogs \footnote{Data are available from\hfill\\
{\tt https://lgarrison.github.io/AbacusCosmos/}}
 from the Abacus project  \citep{Garrison_2018,Garrison_2019}.
A high-resolution simulation was 
run to produce halos with a relatively large number of particles (i.e., $N \sim 10^5-10^6$). 
This simulation has a total of   $700^3$ particle in a box of side $L = 200$ Mpc/h.
The  cosmology was a CDM with a cosmological 
constant and neutrinos included in the background expansion.
With these parameters the particle mass is $M=2 \times 10^9 M_\odot$. 
 Halos are identified by means of the Abacus halo finder, called the CompaSO Halo Finder \footnote{This is a hybrid algorithm described in\\ https://abacussummit.readthedocs.io/en/latest/compaso.html}.
The softening length is fixed in proper, that is, not comoving, coordinates, and it is 
chosen to be 
$\varepsilon= 7$ kpc/h 
\footnote{The convention that $a=1$ at $z=0$ was used, so the proper and comoving softening lengths are both equal to 7 kpc/h at $z=0$. }
(Plummer-equivalent, although spline softening was used).
Then   halo catalogs were generated at a few epochs.

It should be emphasized that there is a long-standing discussion in the literature
concerning small-scale resolution effects in cosmological N-body  that is 
still not clarified. Beyond issues of numerical convergence, 
it is important to understand the limits imposed on
the accuracy of results by the use of a finite number of particles to
represent the theoretical continuum density field, and the associated
introduction of a smoothing scale  $\varepsilon$ in the gravitational
force that  imposes a lower limit on the spatial
resolution. The question
of the  suitable value of the ratio 
$\varepsilon/\ell$, where $\ell$ is the
initial  inter-particle distance, 
 has been the subject of long-standing 
controversy (see, e.g., \cite{joyce+syloslabini_2013,syloslabini_2013b,Baushev_2018}
and references therein). 
The use of a softening length that is fixed in physical coordinates rather than 
in comoving coordinates should mitigate the resolution effects, but a more detailed 
study is needed to proof that this is the case. 
Hereafter we are not going to discuss this issue. Rather 
the point of view we adopt in this work is to identify and study the physical properties of the QSS
without investigating the difficult problem of whether 
resolution effects, especially on small scales, have modified 
the QSS with respect to those expected in the proper 
continuum limit.


\section{Properties of the quasi-stationary states} 
\label{sect:scm+halos}
\subsection{Statistical estimators}
Let us call $k_i$, $\phi_i$, and $e_i$ the kinetic, potential, and total (i.e., $e_i=k_i+\phi_i$) energy of the $i$th particle of the
 system of fixed mass $m_i=m$. Let $K$, $W$, and $E=K+W$ the system kinetic, potential, and total mechanical energy, and $M=Nm$ the total system mass. 
In absence of dissipative mechanisms, the  system total energy
$E (t)$
is clearly conserved along the system evolution as well as its total linear and  angular momenta. 
As mentioned above, by monitoring the behavior of these quantities we have global control of the accuracy of the numerical integration.

The quantity 
\be
\label{eq:delta}
\Delta (t) = \frac{1}{\langle e(0) \rangle } 
\sqrt{ \frac{\sum_{i=1}^N (e_i(t) -  e_i(0))^2} {N} } 
\ee
is a measure of the global exchange of the particle energies over the interval from zero and a generic time $t$, in units of the initial average energy per particle 
\be
\label{e0}
\langle e(0) \rangle = \frac{1}{N}\sum_{i=1}^N e_i(t=0) \;.
\ee

We can consider the estimators 
\be
\label{eq:nr}
\overline{n({\mathbf{r}},t)} = \frac{1}{\Delta V} \sum_{i=1}^{ \Delta N(r)} \delta(\mathbf{r}-\mathbf{r}_i,t),
\ee
\be
\label{eq:kappas}
\overline{k(\mathbf{r},t)} = \frac{1}{\Delta N} \sum_{i=1}^{ \Delta N} k_i(\mathbf{r},t), 
\ee
\be
\label{eq:phis}
\overline{\phi(\mathbf{r},t)} = \frac{1}{\Delta N} \sum_{i=1}^{ \Delta N} \phi_i(\mathbf{r},t), 
\ee
which are volume averages in a sampling volume $\Delta V$ containing $\Delta N$ particles of the number density profile (in Eq.\ref{eq:nr} $\delta(\mathbf{r}-\mathbf{r}_i)$ is the Dirac's delta function), and of the  kinetic and potential energy. 
Other useful statistical indicators are the particle energy distribution, $p(e)$,  and the velocity  distribution
$f(\mathbf{v})$.

The description of the QSS arising from a non-collisional dynamics can be approached in terms of the self-consistent Vlasov-Poisson system of equations \citep{Binney_Tremaine_2008}.
When the long-range force is gravity and specified to stellar dynamics, the Vlasov equation turns in the Jeans equation \citep{Jea1915,Binney_Tremaine_2008}. 
In spherical symmetry, the Jeans equilibrium implies that the function

\be
\label{psi}
\psi(r)= - \dfrac{
\dfrac{\langle v_r^2  (r) \rangle}{\rho(r)} \dfrac{d \rho(r)}{dr}  + 
\dfrac{d \langle v_r^2  (r) \rangle }{dr}  +
\dfrac{2 \beta(r) {\langle v_r^2  (r) \rangle} }{r}  }
{\dfrac{d  \phi(r) }{dr} }
\approx 1 
\ee
where, in Eq. \ref{psi}, $\rho(r)$ is the mass density, $v_r(r)/v_t(r)$ 
the radial/tangential velocity, and 
\[\beta (r)= 1 - \frac{\langle v_t(r)^2\rangle}{2\langle v_r(r)^2\rangle}
\]
the anisotropy parameter such that $\beta=0$ for isotropic orbits and 
$\beta=1$ for radial orbits.


\subsection{Power-law profiles of quasi-stationary states}
Numerical simulations show that the QSS formed by 
the collapse of an isolated, cold, and initally uniform spherical over-density has a density profile of the type 
\citep{vanalbada_1982,aarseth_etal_1988,jmsl_2009, syloslabini_2012}
 \be
\label{vr} 
n_{vr} (r) = \frac{n_0}{1+\left(\frac{r}{r_0}\right)^4}\;,
\ee
where $n_0,r_0$ are two parameters that depend on the specific case under study.

On the other hand, 
 in the context of cosmological simulations in the CDM scenario, 
 a universal  density 
profile nicely fits  the dark matter structures in the highly nonlinear regime, 
the so-called halos. This fitting formula is the Navarro-Frenk-White 
(NFW) profile \citep{NFW_1997,Taylor_2001,Navarro_2004}, that is,
\be
\label{nfw} 
n (r) = \frac{n_0}
{\left(\frac{r}{r_0}\right) \left(1+\frac{r}{r_0}\right)^2} \;.
\ee
The main difference of the profile 
in Eq.\ref{nfw} stands in its cuspy behavior and in its shallower decay at large distances ($r\gg r_0$).

It should be noted that the dynamical processes underlining 
the formation of the profiles in Eq.\ref{vr} and Eq.\ref{nfw} are different. Indeed, in  CDM models the clustering proceeds bottom-up through the subsequent merger of structures into larger structures. 
This happens because 
 initial density fluctuation fields  are 
characterized by long-range correlations. The correlation function decays as $\sim r^{-1}$ in the  
range of scales relevant for cosmological structure formation, 
and, correspondingly, the power spectrum grows as $\sim k^2$ (where $k=2\pi /r$) \citep{Peebles_1983}. 
Although halos are commonly considered as the building blocks of nonlinear
structures formed in a cosmological context, a full theoretical 
understanding of their properties is still lacking (see, e.g., 
\cite{theis+spurzem_1999,Binney+Knebe_2001,diemand_etal_2004,levin_etal_2008}).

With regard to the density profile in Eq.\ref{vr}, 
reached as QSS of an isolated monolithic 
collapse, we show at virialization, the orbital distribution is radially biased, which implies a {\it non-isotropic} velocity distribution \citep{syloslabini_2013}. In what follows we 
derive a similar conclusion using the Jeans equation; further 
we show that this analysis can shed light on the 
 more general case of the density profile in Eq.\ref{nfw}, 
which is not characterized by a single exponent.

Upon the assumption of velocity isotropy many solutions for the distribution function (DF) of given spherically symmetric density laws are found, such as the so-called $\gamma$-model family \citep{Dehnen_1993,ASCD2014}. Nevertheless, 
correct modeling of {\it non-isotropic} (in velocity space) systems, such as those coming from both isolated \citep{syloslabini_2013} and non-isolated \citep{Hansen_2006} $N$-body experiments,
remains an open problem.

\subsection{Case of a uniform and isolated spherical over-density}

The  collapse of an {\it initially uniform} sphere is a  
paradigmatic case  
investigated numerically by a large number of authors 
\citep{henon_1964,vanalbada_1982,aarseth_etal_1988,aguilar+merritt_1990,theis+spurzem_1999,boily_etal_2002,roy+perez_2004,boily+athanassoula_2006, barnes_etal_2009, jmsl_2009, syloslabini_2012, worrakitpoonpon_2014, merritt+aguilar_1985, aguilar+merritt_1990, theis+spurzem_1999,syloslabini_2013, SylosLabini+Benhaiem+Joyce_2015}. 
The specific key role played by density fluctuations during the collapse
has been studied by, for example, \cite{aarseth_etal_1988} and \cite{SpeRCD2017},
while the mechanism of the particle energy change was firstly discussed by \cite{jmsl_2009}.  
We are now going to consider the properties of the QSS,
that are formed after the virialization, in particular the differentiation between core and halo 
\footnote{Unless specified all distances are expressed in kpc and all times in Gyr. The velocities are measured in km/s.}.

\begin{figure}
\includegraphics[angle=0,width = 3.0in]{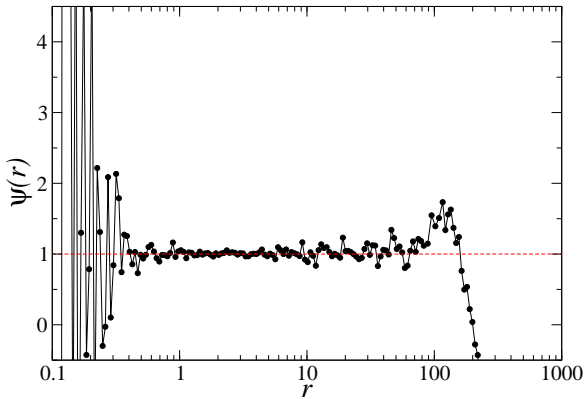}
\caption{Behavior of the function $\psi(r)$ defined in Eq.\ref{psi} at $t=9$ Gyr  in the case of the initially uniform sphere, where $N=10^6$. 
At small distances the deviation from $\psi=1$ is due to sparse sampling fluctuations while at large distances, (i.e., $r>100$ kpc) the deviation is due to the out-of-equilibrium nature of the system.}
\label{psi_1e6} 
\end{figure}

The QSS is in equilibrium and indeed  
Eq.\ref{psi} is satisfied (see Fig.\ref{psi_1e6}).
The signal is noisy at small distances because 
the number of particles in shells is small (i.e., $N<10^2$), and at larger distances (i.e., $r>100$ kpc)
there is a clear deviation because the particles have positive energy.
The behavior of $\psi(r)$ for this system represents a useful reference for the analysis of
the more complex situations presented in what follows.

\begin{figure}
\includegraphics[angle=0,width = 3.0in]{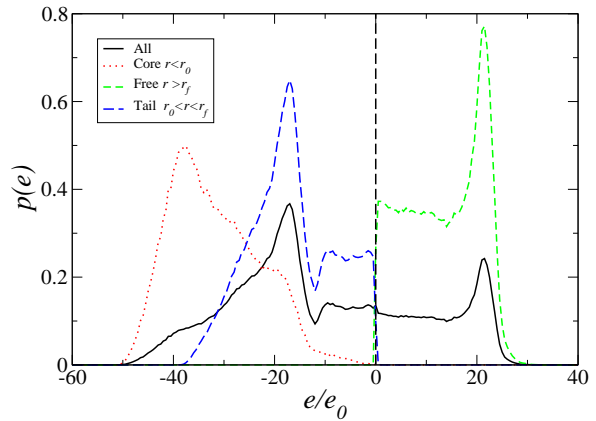}
\caption{
Particle energy distribution of the QSS at $t=9$ Gyr
in units of $e_0$ (see Eq.\ref{e0}).
The system after the collapse is made of three components:
two form the QSS (tail and core) and the third, where $e>0$, is made of ``free'' particles. The core is  comprised of particles 
with radial distance $r<r_0$; the free particles
have positive energy or $r>r_f$, where $r_f = r_f(t)$ must be estimated from the numerical data.
}
\label{pe_1e6} 
\end{figure}

%
Let us now consider the core and the halo of the QSS separately. 
The core is defined as the region within the length scale $r_0$ 
found by fitting the density profile with Eq.\ref{vr}.
Figure \ref{pe_1e6} shows the normalized particle energy distribution 
 of the QSS at $t=9$ Gyr (i.e., at a time much longer than 
$\tau_{ff} \approx 1.5$ Gyr).
In particular, the three main components of the system after the collapse are highlighted:
the first two constitute the actual QSS, namely the core (i.e., $r<r_0$), and the outermost bound particles that 
forms the halo ($r>r_0$ and $e<0$),
while the remaining component is comprised of free particles.

A gas under steady-state conditions at a temperature $T$  
immersed in a conservative force field is characterized by a distribution 
function that differs from the Maxwell-Boltzmann  (MB) distribution
by the exponential factor $\exp(- \Phi (\vec{r})/kT) $, 
where in this case the temperature can be defined through the 
particle velocity dispersion. 
In this situation, the equilibrium distribution function for this
 case is
written as \be
f(\vec{v},\vec{r})   =
 n_0 \left( \frac{m}{2\pi k T} \right)^{3/2} \exp\left(- \frac{mv^2/2+\Phi (\vec{r})}{kT}, \right) \;. 
 \ee
Consequently, the number density for a system described by this distribution function is
 given by
 \be
 \label{densitycore} 
  n(\vec{r})  = n_0 \exp(- \Phi (\vec{r})/kT). 
  \ee

\begin{figure}
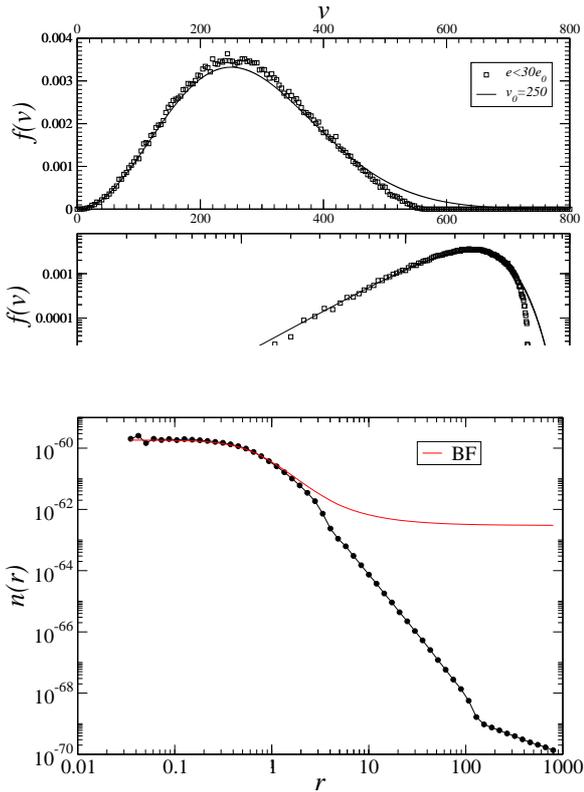

\includegraphics[angle=0,width = 3.in]{Fig3a.eps}
\includegraphics[angle=0,width = 3.in]{Fig3b.eps}
\caption{
Top   panel: 
The velocity distribution function in the core 
by applying a selection i in energy 
and the  corresponding best fit with 
a MB distribution with $v_0=250$ km/s 
in linear and (middle panel) bi-logarithmic scale.
Bottom  panel: Measured density profile together with the 
Boltzmann factor 
(see Eq.\ref{densitycore}).
}
\label{UnifSph_fv} 
\end{figure}

The velocity distribution function in the core (see the 
top and middle panels of Fig.\ref{UnifSph_fv})
is well approximated by a MB distribution.
We carried out the fit by defining the core in two different ways: i) all particles with $r<r_0$,
where $r_0$ was estimated from the best fit of the density profile; 
and ii) by considering an energy threshold, that is,  $e/e_0 <-20$,
where $e/e_0=-20$ corresponds to the inner peak of $p(e)$ and
$e_0$ is defined in Eq.\ref{e0}).
In the latter case the fit is better than in the former.
We interpret this as a consequence of
the energy threshold selecting the particles in the inner
core better than the distance cut because in that case 
particles that have higher
energy, and thus belonging 
to the halo at a subsequent time, can be confused 
with the core particles. 
The temperature 
$T$ can be thought to be an effective temperature related to  
an isotropic and scale-independent velocity dispersion, that is, it does not 
represent a real equilibrium 
thermodynamical temperature.
Fig.\ref{UnifSph_fv} (bottom panel) shows the almost flat 
density distribution in the central region where, additionally, 
the velocity distribution is isotropic. 

Thermal equilibrium is reached in the core driven by two-body collisions. The order of magnitude of the timescale of 
collisional relaxation is \citep{Binney_Tremaine_2008}
given by\be 
\tau_{2b} \approx \frac{N}{\log N} \tau_{dyn} \approx \frac{N}{\log N} \sqrt{\frac{\rho} {\rho_0} } \tau_{ff} 
,\ee
where $\tau_{ff}$ is the free-fall time of the system  (that 
has initial density $\rho$; see Eq.\ref{tauff}) and 
$\tau_{dyn} \sim  {(G\rho_0)}^{-1/2}$ is the dynamical time of the core with density $\rho_0 \gg \rho$.
In the core (i.e., for $r<r_0$), we find that
${\rho}/{\rho_0} \sim 10^{-5}$ and $ {N}/{\log(N)} \sim 10^2-10^3$
($\log$ is the decimal logarithm) and thus $\tau_{2b} \approx \tau_{ff}$: two-body relaxation is efficient enough to establish
thermal equilibrium in the core in a timescale of order of $\tau_{ff}$.
 {We note that an approximate thermal equilibrium is reached in the short timescale corresponding to the global collapse of the system
 $ \tau_{dyn}$ and that on a much longer timescale, driven by two-body encounters, eventually the QSS that emerges from
 the violent relaxation process undergoes a gravothermal collapse.}

\begin{figure}
\includegraphics[angle=0,width = 3.0in]{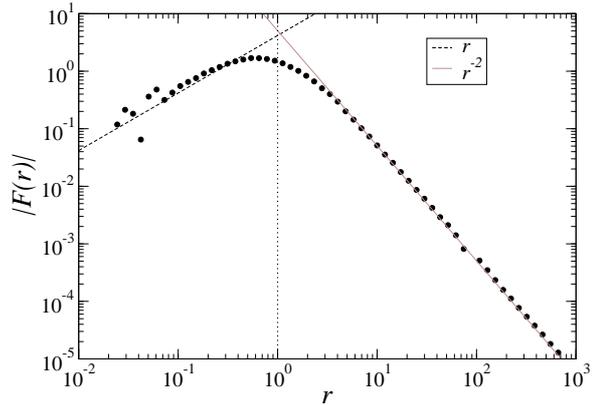}
\caption{
Absolute value of the force 
as a function of scale in the QSS.}
\label{force_1e6} 
\end{figure}

The power-law fit to the density profile is very well defined for $r \geq 4 r_0$. 
Such a region contains 
most of the mass  of the system
and it is  surrounded  by  
a lower density region of bound particles, 
still  spherically symmetric  distributed, 
in which the density displays a power-law decay and whose velocity distribution is radially biased. 
Given these conditions, we aim to find the relation between 
the exponent $\alpha$ of the power-law fit to the density profile (i.e., $\rho(r) \sim r^{-\alpha}$) 
and the anisotropy parameter $\beta(r)$ given that the Jeans equation 
(Eq.\ref{psi}) is satisfied.  
We note that, 
under the hypotheses mentioned above, 
the gravitational potential decays, for $r >r_0$, as 
$\phi(r) \sim -{GM_0}/{r}$ 
--- thus corresponding to a force that decays as $r^{-2}$ 
 (see Fig.\ref{force_1e6}).

\begin{figure}
\includegraphics[angle=0,width = 3.3in]{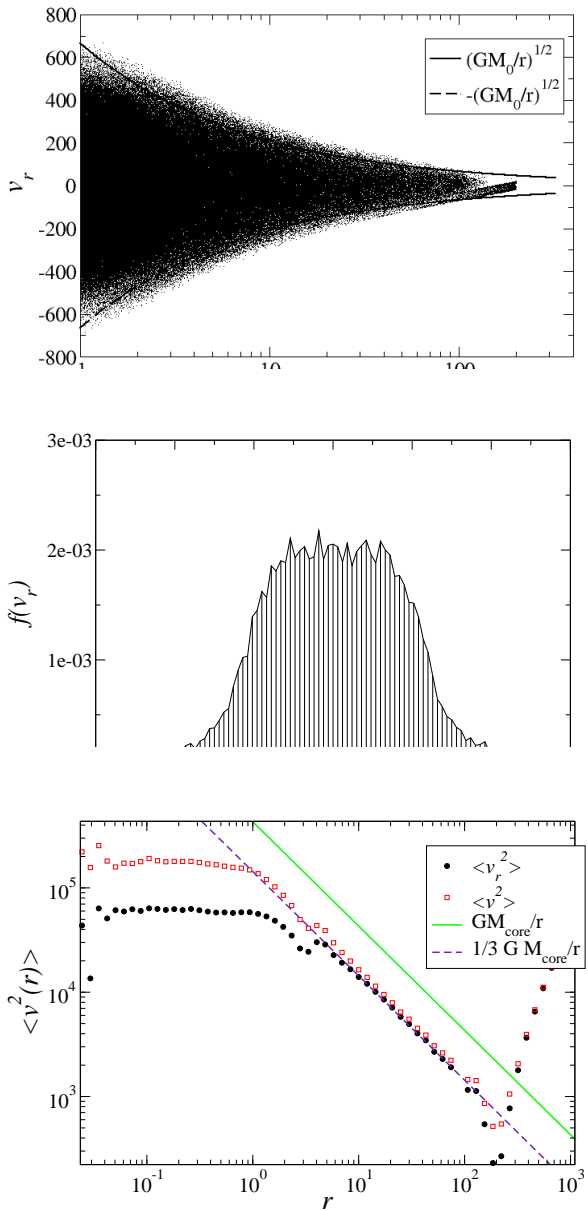}
\includegraphics[angle=0,width = 3.in]{Fig5b.eps}
\includegraphics[angle=0,width = 3.in]{Fig5c.eps}
\caption{
Top panel: The radial component of the velocity 
$v_r$ as function of the  distance for particles in the tail.
Middle panel: Example of the 
velocity distribution $f(v_r)$ in a tail 
shell.
Bottom panel:  $\langle v_r^2(r)\rangle $ and 
$\langle v_r^2(r) \rangle$ as function of the radial distance}
\label{vr_1e6} 
\end{figure}

In this external zone, owing to low density, the self-interaction between particles can be neglected 
so that the  maximum speed of a particle at distance $r$ is the local escape velocity (i.e., $v_r^M(r)  =  \sqrt{ {GM_0}/{r} }$; see the top panel of Fig.\ref{vr_1e6}).

By assuming that the probability distribution function (PDF) of $v_r(r)$ is uniform 
in the range $[-v_r^M(r),v_r^M(r)]$ , a situation that occurs if the system 
is virialized (see the middle panel of Fig.\ref{vr_1e6}), 
{we find}
\be
\label{v2rm}
\overline{v_r^2(r)} = 
\int_{-v_r^M(r)}^{v_r^M(r)} p(v_r) v_r^2 dv_r = 
\frac{1}{3} \frac{G M_0}{r} 
.\ee
The bottom panel of Fig.\ref{vr_1e6} shows that Eq.\ref{v2rm}  well approximates the measured behaviors.
From Eq.\ref{psi} and Eq.\ref{v2rm} we find that  for $r>r_0$ 
\be
\frac{1}{\rho(r)} \frac{d \rho(r)}{dr}
=
\frac{1}{r} 
\left(
\frac{\phi_0}{\overline{v_r^2(0)}} + 1 -2 \beta(r)  
\right) = 
\frac{-2 (1 + 
\beta(r)  
)}{r} \;.
 \; 
\ee
If we take $\beta(r)=1$ 
we find 
\be
\rho(r) \sim  \frac{\rho_0}{r^4} \;,
\ee
which well approximates the power-law tail 
observed in numerical simulations (see
Eq. \ref{vr}).

{In these  same 
approximations we find, for $r \gg r_0$ }
\be
\label{alpha}
 \alpha \approx   2(1+\beta)  \rightarrow  4 \; \; \mbox{for}  \; \;  \beta  \rightarrow  1\;.
\ee
However, we urge caution in extrapolating Eq.\ref{alpha}
for any value of $\beta,\alpha$. In 
  general the situation is more complicated as neither Eq.\ref{v2rm} 
 nor $|\phi| \sim 1/r$ is satisfied when the density decays slower than $r^{-4}$ and 
we should consider Eq.\ref{psi} instead of Eq.\ref{alpha}
and thus $\alpha$ is expected to have a nontrivial dependence 
on $\beta$ and on the whole mass distribution.

\begin{figure}
\includegraphics[angle=0,width = 3.in]{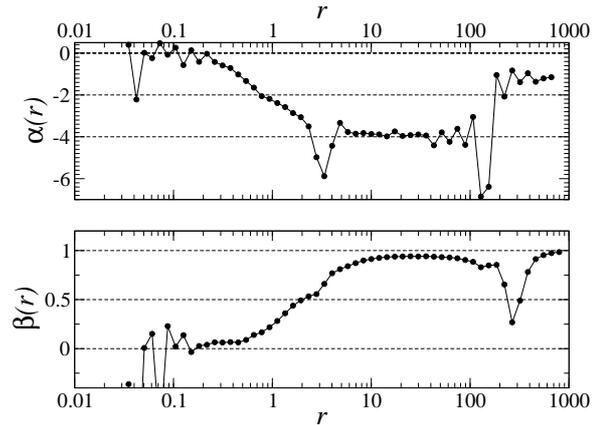}
\caption{
Top panel: Radial behavior of the exponent 
$\alpha(r)$ of the density profile.
Bottom panel: Radial behavior of the anisotropy parameter $\beta(r)$.}
\label{alpha_beta_UnifSph} 
\end{figure}

In summary, in the case of a violent relaxation of a 
isolated, cold, spherically symmetric, uniform mass distribution 
we obtained the limiting behaviors (see Fig.\ref{alpha_beta_UnifSph})
\bea
&&
\alpha \rightarrow 0 \;\; \mbox{for}  \;\; \beta \rightarrow 0 \;\; r \le r_0
\\ \nonumber &&
\alpha \rightarrow 2(1+\beta) \;\; \mbox{for}  \;\; \beta \rightarrow 1   \;\; r \gg r_0\;.
\eea

%

\subsection{Isolated spherical  over-densities with non-Poissonian fluctuations}

As discussed  Sect.\ref{sect:simulations}
the key parameter 
of the second family of IC
is $\gamma$ (see Eq.\ref{gamma});
when $\gamma=10$ the number of centers
is only ten times less than the number of particles,
and thus fluctuations are slightly greater than in the purely Poissonian case.
On the contrary, when $\gamma \approx 10^5$
the IC consist of subclumps that  have collapse timescales 
shorter than that of the system as a whole. In this situation, 
subclumps collapse almost independently from each other 
and then the different substructures merge. A separation of spatial and temporal scales occurs only when the IC is highly inhomogeneous with few centers
The intermediate range for $\gamma$, $10<\gamma<10^5$, is the most interesting to study. 

\begin{figure}
\includegraphics[angle=0,width = 3.in]{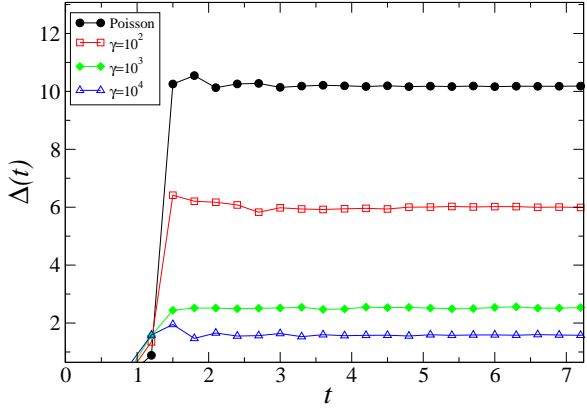}
\caption{Behavior with time of the quantity $\Delta(t)$ 
(see Eq.\ref{eq:delta}) in simulations of 
isolated spherical over-densities with non-Poissonian fluctuations with different values of $\gamma$.
The behavior for the case of the initially
Poissonian  spherical over-densities (black) is shown for comparison.
}
\label{fig:delta_sub} 
\end{figure}

\begin{figure}
\includegraphics[angle=0,width = 3.in]{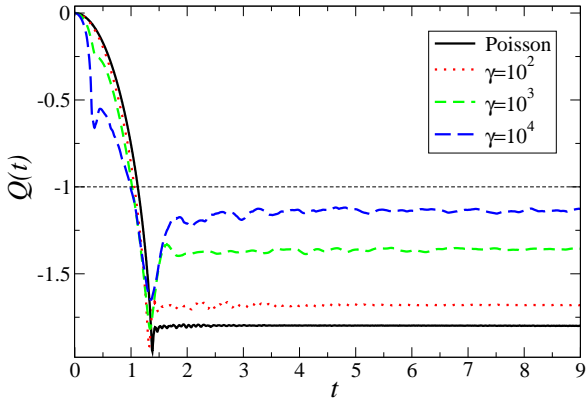}
\caption{Time evolution of the virial ratio $Q$  
in simulations of isolated spherical  over-densities with non-Poissonian fluctuations
with different $\gamma$.
The case of the initially Poissonian spherical over-density is also shown for comparison (in black).
}
\label{fig:b_sub} 
\end{figure}

\begin{figure}
\includegraphics[angle=0,width = 3.in]{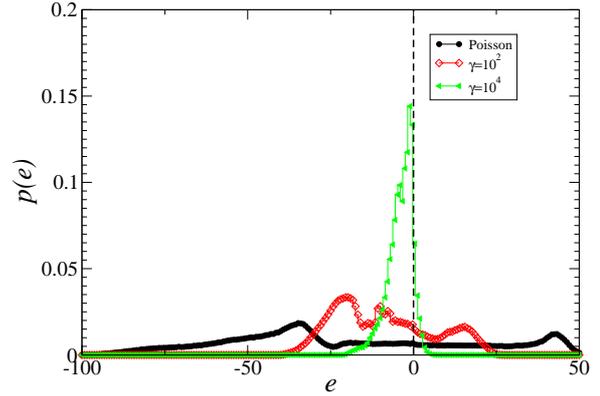}
\caption{Asymptotic particle energy distribution 
in the two non-Poissonian  simulations.  
The case of a Poissonian IC is also shown for comparison. 
}
\label{fig:pe_sub} 
\end{figure}

Figure \ref{fig:delta_sub} shows the behavior of the quantity $\Delta(t)$,
defined by Eq.\ref{eq:delta},
which measures the amount of energy exchanged 
among system particles. A clear tread shows that 
the  more uniform the IC the larger is the energy variation.
This trend is in line with the Poissonian case,
where the larger $N$ is, the smaller the initial fluctuations over the mean and the larger the variation of $\Delta(t)$ \citep{jmsl_2009}. 
Consequently, the asymptotic value of the virial ratio becomes closer to $-1$ as the energy exchange gets smaller,
and thus the amount of particles that have been ejected from the system after the collapse is smaller 
(see Fig.\ref{fig:b_sub}).
The reflection of this situation can be clearly seen in the asymptotic shape of the particle 
energy distribution (see Fig.\ref{fig:pe_sub}). The more clustered the initial distribution is, the softer the collapse
and the less spread $p(e)$ after the collapse. 

\begin{figure}
\includegraphics[angle=0,width = 3.0in]{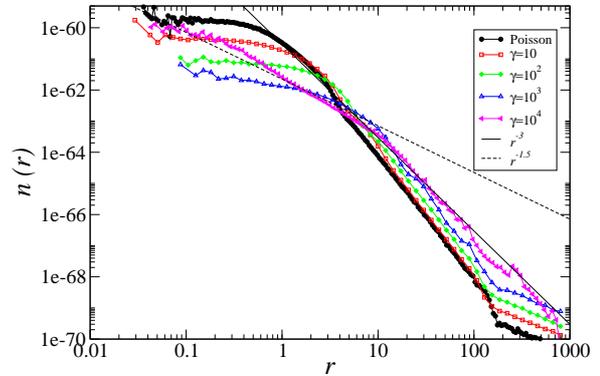}
\caption{Density profile  in the three simulations of 
isolated spherical  over-densities with non-Poissonian fluctuations with different $\gamma$.
The behavior for the case of the initially
Poissonian  spherical  over-density is also shown for comparison (in black).}
\label{fig:nr_sub} 
\end{figure}

The density profiles of the QSS in the various simulations are shown by Fig.\ref{fig:nr_sub}. 
While there is a clear change of slope in all cases between the inner core and the outer 
regions, the softer the collapse and the less marked is such a change. 
That is, while for the uniform case there is a clear change from $n(r) \sim$ const.
in the inner core to $n(r) \sim r^{-4}$ in the outer regions, when the initial 
fluctuations are large enough (i.e., $\gamma \approx 10^2-10^4$), 
then the density profile in the inner core is closer to 
$n(r) \sim r^{-1}$ and in the outer region to $n(r) \sim r^{-3}$. 

The QSS formed are close to the Jeans equilibrium in all cases. We find that $ \psi(r) \approx 1$ in an intermediate 
range of scales between the inner regions, where shot noise fluctuations
are predominant, and the outermost regions, where particles have 
positive energy.

To clarify the statistical 
and dynamical properties of the QSS 
we focus on the case in which initial fluctuations are large but the number of subclusters is still
large enough so that they have a substantial overlap and thus there is not a separation 
of length and timescales in the collapsing phases of the whole structure and its substructures. 
We thus focus on the case 
 $\gamma=10^4$.  
 
\begin{figure}
\includegraphics[angle=0,width = 3.0in]{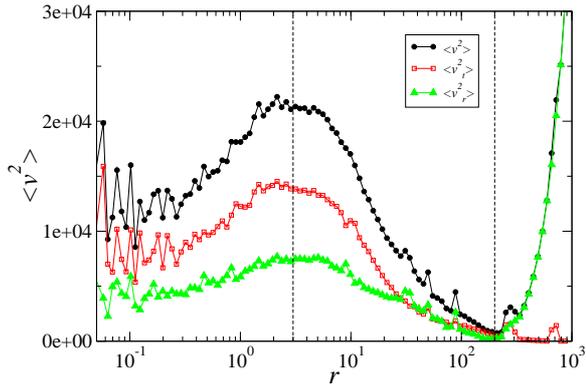}
\caption{Velocity dispersion (total, transverse, and radial) 
as a function of distance in the asymptotic QSS for  a
the case of a  very clustered IC, i.e.,  $\gamma=10^4$.}
\label{fig:v2r_sub2} 
\end{figure}

\begin{figure}
\includegraphics[angle=0,width = 3.0in]{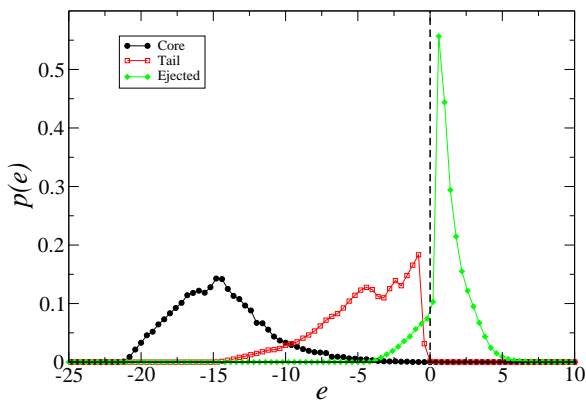}
\caption{Particle energy distribution 
in the asymptotic QSS for the case of a  very clustered IC, i.e., $\gamma=10^4$}
\label{fig:pe_sub2} 
\end{figure}

 Fig.\ref{fig:v2r_sub2} shows the velocity dispersions  
as a function of distance to the center in the asymptotic QSS. 
We can identify three different regimes that correspond to the following: 
\begin{itemize}
\item (i) an inner region, in which the 
dispersion grows slightly inward in an almost isotropic 
manner (i.e., 
$\langle v_t^2 \rangle \approx 2 \langle v_r^2 \rangle)$,
corresponding to $\beta(r)\approx 0$, 
\item (ii) a tail, in which the dispersion (in all components) decreases as a function of the radial distance and 
$\langle v_r^2 \rangle > \langle v_t^2 \rangle$,
\item (iii) an outermost region, in which $\langle v_t^2 \rangle \approx 0$, that is, where highly energetic particles move on quasi-radial orbits. 
\end{itemize}
Figure\ref{fig:pe_sub2}  shows the particle 
energy distribution.
Particles in the core are selected as those having
$r < r_0$, where in this case $r_0$ approximately coincides 
with the peak of $\langle v^2(r) \rangle$
(see Fig.\ref{fig:v2r_sub2}) 
and 
the ejected particles 
have $r>r_f$, where $r_f$ is estimated
to be the (time-dependent) scale at which $\langle v^2(r) \rangle 
\approx \langle v_r^2(r) \rangle$. 
The core is populated by the most bound particles, the tail is made by particles with slightly negative energy, while  particles in the outermost region have $e>0$. 

\begin{figure}
\includegraphics[angle=0,width = 3.0in]{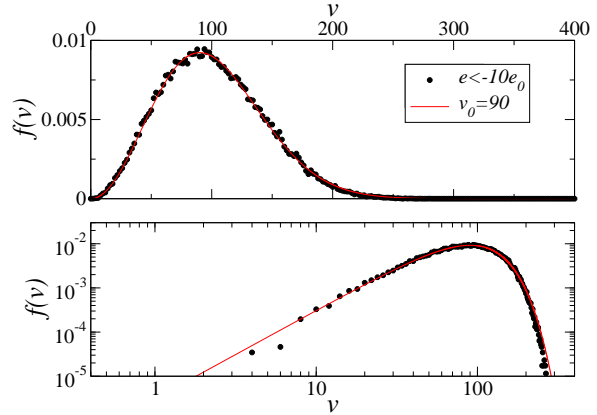}
\caption{Velocity distribution of the inner regions 
of the asymptotic QSS for  a
the case of a  very clustered IC, i.e.,  $\gamma=10^4$.
The selection was performed in energy (i.e., $e<10 e_0$);
the  
best-fitting Maxwell-Boltzmann distribution  is reported.}
\label{fig:pv_sub2} 
\end{figure}

Figure\ref{fig:pv_sub2} shows 
the velocity distribution of the particles in
the core region. As for the case 
of the Poissonian IC, a MB distribution represents a good fit, clearly at a much lower temperature (i.e., velocity dispersion)
than for the uniform case. 
We selected the inner regions 
in two ways: by considering a limit 
in radial distance (i.e., $r<r_0$) and
a limit in energy ($e<- 10\, e_0$). In the latter case the MB distribution fit better interpolates the data.
In the inner core, thermal equilibrium is reached driven 
by tow-body collisions even in this case. 

\begin{figure}
\includegraphics[angle=0,width = 3.0in]{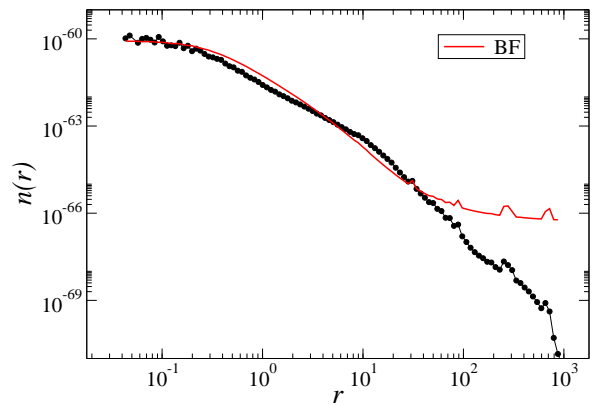}
\caption{Density profile in the inner core  
of the asymptotic QSS for
the case of a  very clustered IC, i.e.,  $\gamma=10^4$; the 
best-fitting with Eq.\ref{densitycore}
is also shown.} 
\label{fig:nr_theo_sub2} 
\end{figure}

\begin{figure}
\includegraphics[angle=0,width = 3.0in]{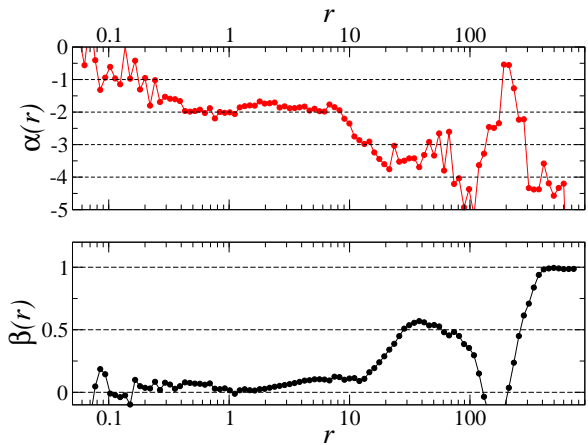}
\caption{
Top panel: Behavior of the exponent of the density profile $\alpha(r)$ 
of a QSS with $\gamma=10^4$.
Bottom panel: Behavior of the anisotropy parameter $\beta(r)$  as a function 
of  the distance from the center. 
} 
\label{fig:alphabeta_sub2} 
\end{figure}

\begin{figure}
\includegraphics[angle=0,width = 3.0in]{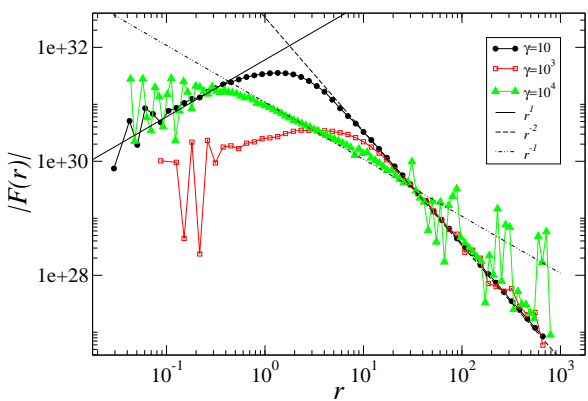}
\caption{
Absolute value of the gravitational force 
as a function of  function of scale in the QSS
in simulations with different values of $\gamma$.
}
\label{force_sub_1e6} 
\end{figure}

Figure \ref{fig:nr_theo_sub2}  shows the 
density profile in the inner region for 
$\gamma=10^4$. 
Even in this case, the density distribution is well described by
Eq.\ref{densitycore}. 
Figure\ref{fig:alphabeta_sub2} shows the behavior of the
anisotropy parameter $\beta(r)$ (bottom panel)  and that of the exponent 
of the density profile $\alpha(r)$ (top panel) as functions 
of  the distance from the center. We note that, as in the case of the 
cold uniform spherical over-density, $\beta(r) \rightarrow 0$ in the core 
and   $\beta(r) \rightarrow 1$ in the outermost region; 
correspondingly the exponent of the density profile $\alpha  \rightarrow 0$ in
the core and $\alpha  \rightarrow -4$ in the tail. Beyond
these two limiting cases it is not possible to obtain an
analytical expression of $\alpha(\beta)$ for the general case. 
The actual mass distribution is more spread than in the simplest model, where $\rho \sim$ const. in the inner region 
and $\rho \sim r^{-4}$ in the outer tail. 
This is quantitatively illustrated by 
the behavior of the 
gravitational force in models with different values of
$\gamma$ (see Fig.\ref{force_sub_1e6});
at short distances from the center the 
linear growth (implied by a constant matter density) is clear 
only when $\gamma < 10^3$. For large $\gamma$, for instance $\gamma=10^4$, the force has a short range of radial growth to decay after as $r^{-1}$ in a 
intermediate range of distance scales. 


\subsection{Cosmological halos}

 As mentioned in Sect.\ref{sect:simulations} we also considered a set of data extracted
 from the Abacus simulations \citep{Garrison_2018,Garrison_2019} representing the halos typical of cosmological simulations. 
 Their shape is typically ellipsoidal  and is characterized
by  several substructures. 
Nevertheless, we treat these systems in spherical symmetry,  
as the ratio between the axes is close to one, 
and we 
compute the center as the minimum of the potential energy.
A certain degree of arbitrariness in the definition of the outermost cutoff of a halo is present.
In this work, we just consider the outputs of the Abacus halo finder keeping in mind that
faraway low density particles with high energy may not be included because of a selection effect.

In what follows, we report results for the
 three more massive halos,  H49850, H40661, and H965,
which contain, respectively, $\sim(8, 4, 3)\times 10^5$ particles. We checked that when considering 
smaller halos the results do not 
qualitatively change but the statistical estimators are  noisier.

\begin{figure}
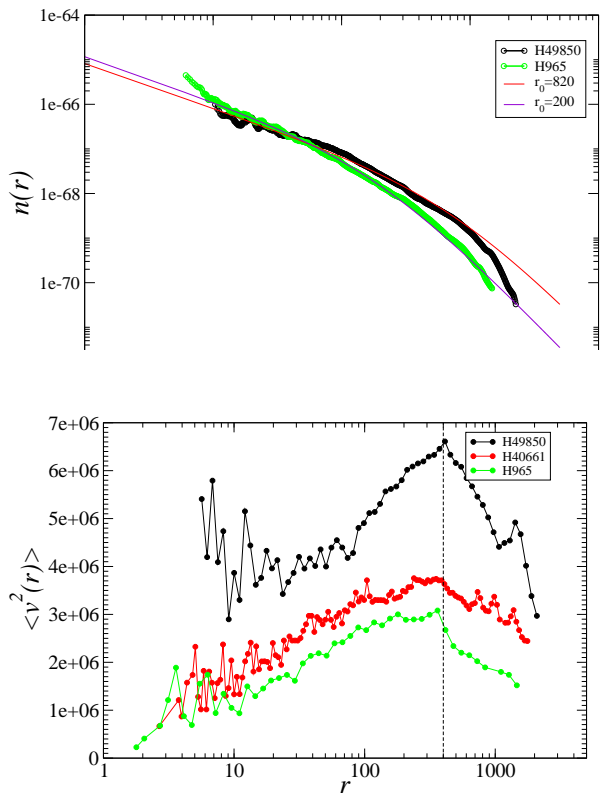

\includegraphics[angle=0,width = 3.2in]{Fig17a.eps}
\includegraphics[angle=0,width = 3.in]{Fig17b.eps}
\caption{
Top  panel: Density profile in two of 
most massive  halos. The two solid lines represent the best fit 
for H49850 and H965, respectively, with a NFW profile. 
 Bottom panel: The velocity dispersion  $\langle  v^2(r) \rangle$ 
 for the same halos together with H40661.
}
\label{fig:nr_abacus} 
\end{figure}

The density profiles of two  halos are shown in the top panel of Fig.\ref{fig:nr_abacus}. 
The density profile slope changes from $\alpha \approx -1$ in the inner region to $\alpha=-2$
in the outer region of the system. The  NFW profile (see Eq.\ref{nfw})
provides good fits of the behaviors but in the outermost part of the tail,
where there is some arbitrariness in the definition of the particle memberships.
The radial behavior of the average square velocity resembles  that observed in isolated spherical collapse models with non-Poissonian fluctuations
(see Fig.\ref{fig:v2r_sub2}): indeed, $\langle  v^2(r) \rangle$ 
grows with distance reaching 
a maximum at $\sim r_0$ and then decays
at large distances.
The radial and transverse velocity dispersion (not plotted) display a 
similar behavior.
The radial scale $r_0$ roughly separates the two regimes. 

\begin{figure}
\includegraphics[angle=0,width = 3.0in]{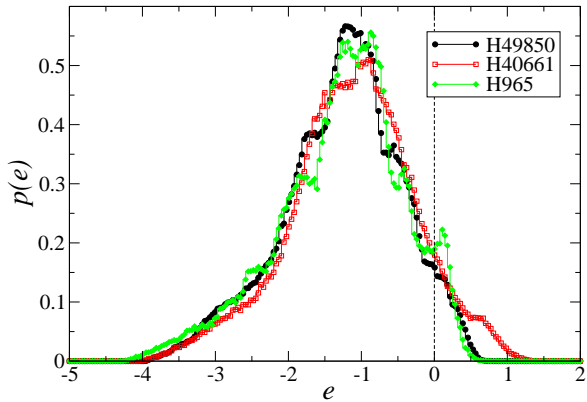}
\caption{
Particle energy distribution in the three largest  Abacus halos.
In this case the energy is normalized to $e_0=Wm/M$, where 
$W$ is the    total gravitational 
potential energy of the system at $z=0$   (i.e., not the initial one), 
$M$ is 
its mass, and   $m$ the particle mass.}
\label{fig:pe_abacus} 
\end{figure}

The particle energy distribution in shown in Fig.\ref{fig:pe_abacus}; a small fraction of the
particles have positive energy in all the three cases. This fraction clearly depends on the manner in which the external part of
the halos has been selected. 
Differently from Fig.\ref{pe_1e6} and Fig.\ref{fig:pe_sub} 
in this case the energy is normalized to $e_0=Wm/M$, where $W$ is the  
gravitational potential energy of the system
at redshift $z=0$ (i.e., not the initial), $M$ its mass, 
and $m$ the particle mass: this is clearly much larger (in absolute value) 
than the initial one. 
We treat each halo as being isolated; this is
clearly an approximation that works better  
when the halo density contrast is larger.
The overall 
shape of these $p(e)$ is very similar 
to those obtained in the case of isolated spherical 
collapse models with non-Poissonian fluctuations and large $\gamma$.
(see Fig.\ref{fig:pe_sub}).

\begin{figure}
\includegraphics[angle=0,width = 3.0in]{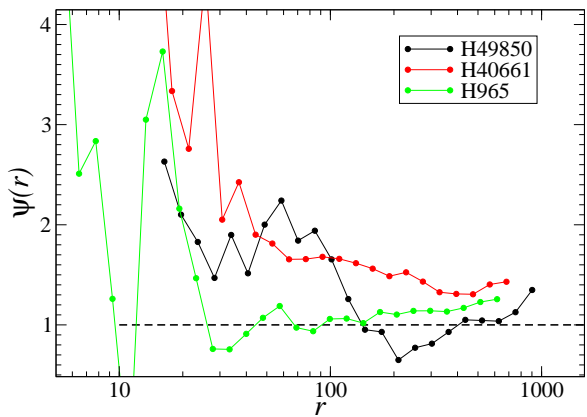}
\caption{
Behavior of the parameter $\psi(r)$ (see Eq.\ref{psi}) for the three largest  Abacus halos.}
\label{fig:psi_abacus} 
\end{figure}

The halos are close to an equilibrium condition as described by the Jeans equation
 (see Eq.\ref{psi}), that is,  $\psi(r) \approx 1$ (see Fig.\ref{fig:psi_abacus}).
 At small distances $\psi(r)$ has  shot-noise fluctuations while the outermost region
 of the tail is out of equilibrium as particles have positive energy. 
 In the intermediate region $\psi(r)$ presents larger fluctuations
 than for the case of the uniform sphere (Fig.\ref{psi_1e6}).
 This is probably results from the 
 presence of more substructures and the influence of
 neighboring density perturbations because now these 
 over-densities are not isolated as in the previous cases.

\begin{figure}
\includegraphics[angle=-0,width = 3.in]{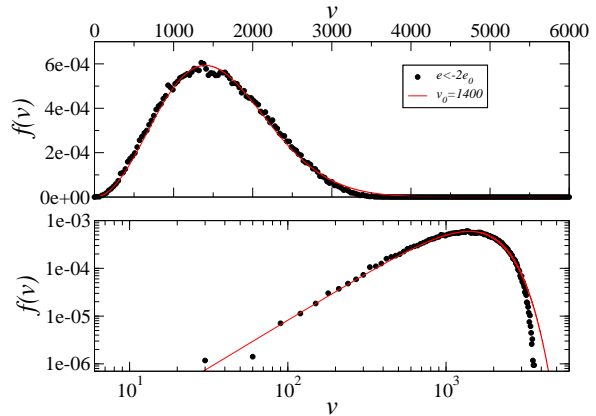}
\caption{
 Velocity distribution in the inner core  for the most massive 
 Abacus halo (i.e., H49850)  
together with the best fit with a MB 
distribution. 
Inner region particles were selected by considering 
an energy cut (i.e., $e < -2 e_0$) .}
\label{fig:fv_abacus} 
\end{figure}

\begin{figure}
\includegraphics[angle=0,width = 3.0in]{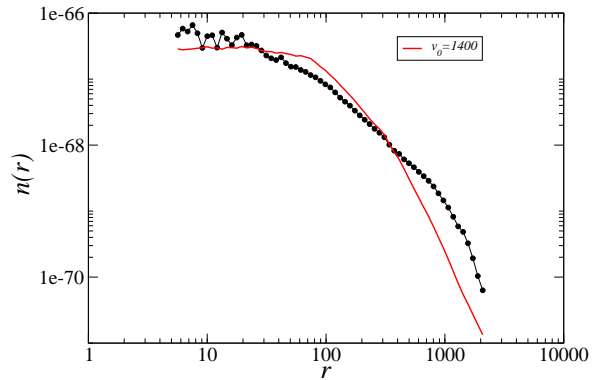}
\caption{Density profile of the  most massive 
 Abacus halo (i.e., H49850).  The 
best-fitting halo
with Eq.\ref{densitycore} is also shown.
 }
\label{fig:BF_abacus} 
\end{figure}

The inner region 
shows the velocity distribution that is well approximated by a 
MB distribution (see Fig.\ref{fig:fv_abacus}).
Inner region particles were selected by considering 
a radial distance cut (i.e., $r<r_0$) 
and, alternatively, an energy cut; in the latter case the MB distribution fits the data better,
 as for the cases of Fig.\ref{UnifSph_fv} and Fig.\ref{fig:pv_sub2}.
If we use Eq.\ref{densitycore} to compute the density profile of the inner region we obtain a fit that is worse than 
in the cases discussed previously (see Fig.\ref{fig:BF_abacus}) but that is still reasonably good.
On the other hand, the fit is particularly rough at large radial distances. This is probably because the halo is
not isolated in these simulations system.

\begin{figure}
\includegraphics[angle=0,width = 3.in]{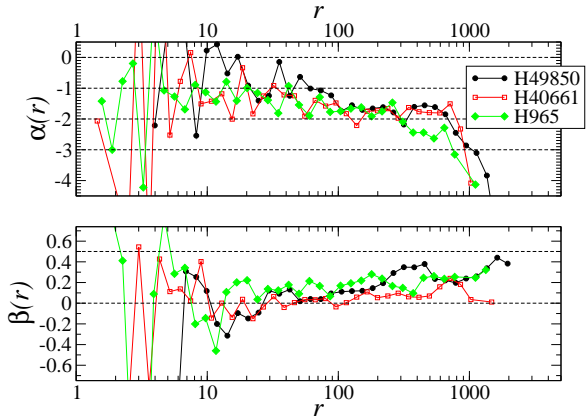}
\caption{
Top panel: Derivative of the density profile for the three largest Abacus halos.
Bottom panel: Anisotropy parameter. 
 }
\label{fig:alpha-beta_abacus} 
\end{figure}

Figure \ref{force_abacus}
shows the behavior of the absolute value of the gravitational force in the three examined massive halos. This value is approximately constant at small radii
and then decays as $\sim r^{-1}$ at large distances,
which is a behavior consistent with the
density profile shown in Fig.\ref{fig:nr_abacus}.
The large fluctuations in the force profile, especially for the case of the halo H49850,
are due to substructures. Finally, the  anisotropy parameter 
 (bottom panel of Fig. \ref{fig:alpha-beta_abacus}) 
is $\beta \approx 0$ in the inner 
zone (where $n(r) \sim r^{-1}$), while $\beta \approx 0.5$ in the outer regions of the system (where $n(r) \sim r^{-2}$).

\subsection{Discussion} 

\begin{figure}
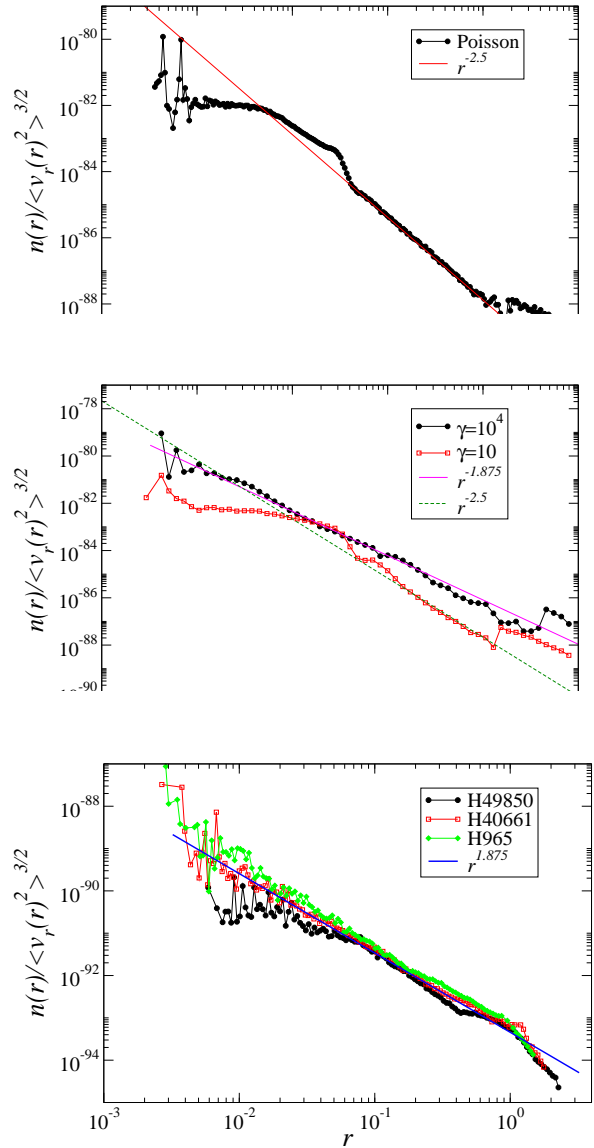

\includegraphics[angle=0,width = 3.in]{Fig24a.eps}
\includegraphics[angle=0,width = 3.in]{Fig24b.eps}
\includegraphics[angle=0,width = 3.in]{Fig24c.eps}
\caption{Coarse-grained phase-space density in the uniform case (top panel),
nonuniform case (middle panel),
and  cosmological halos (bottom panel). }
\label{fig:phsp} 
\end{figure}

In  high-resolution simulations of 
CDM halos \citep{Taylor_2001}  the coarse-grained  
phase-space density decays as 
 \be
 \label{phsp} 
 \frac{\rho(r)}{\langle v^2_r (r)\rangle^{3/2} } \sim \frac{1}{r^\mu} \;,
 \ee
where   $\mu \approx1.875$. 
We compare the results of our experiment with the above behavior, starting from the violent collapse case of a uniform sphere. In this case, combining Eq.\ref{vr} and Eq.\ref{v2rm} we find $\mu=5/2$ in the tail, while in the core $\mu \simeq 0$  (see top panel of Fig. \ref{fig:phsp}).
The QSS established after the collapse of spherical distributions with non-Poissonian fluctuations 
show a behavior that depends on
the parameter $\gamma$: when $\gamma$ is small, and thus the initial fluctuations are 
close to Poissonian, then the slope $\mu$ is similar to that of the Poissonian case. When 
instead the distribution is initially clustered (i.e.,
$\gamma=10^3-10^4$), then the 
 slope is $\mu \approx 1.875$
(see the middle panel of Fig. \ref{fig:phsp}).
  The different behavior observed for the isolated and clustered simulations as a function of $\gamma$ 
  is again good evidence that, by changing such a parameter, the mean-field and collisionless 
  dynamics that drive system to reach a QSS passes from being close to a top-down monolithic collapse
  to a bottom-up hierarchical aggregation process. 
Finally, in the cosmological halos of the Abacus simulations of the previous subsection we find $\mu \approx1.875$, thus very similar
to the isolated case with $\gamma =10^3-10^4$
  (see the bottom panel of Fig.\ref{fig:phsp}).

For the case of cosmological halos, there have been attempts to determine the slope of the density profile $\alpha$
for  spherically symmetric and {\it isotropic} systems that are in Jeans 
equilibrium  and that exhibit  power-law, 
 coarse-grained, phase-space density \citep{Taylor_2001}. It was formally shown that the allowed density slopes $\alpha$ 
 lie in the range $[1,3]$ \citep{Hansen_2004}.
 It was then noticed that
 in halos extracted from cosmological simulations 
 there  is a linear relationship between the density slope and the
anisotropy  parameter \citep{Hansen_2006,Hansen_2006b,Hansen_2006c}; this relationship, however, has a large scattering,  where for 
$\alpha \rightarrow 3$ for 
$\beta \rightarrow 0.5$ 
and 
$\alpha \rightarrow 1$ for 
$\beta \rightarrow 0$.
These trends are similar to what we obtain in examining the Abacus halos,
 but not for case in which the more violent relaxation 
 occurring when a monolithic collapse takes place as  
in this latter case  $\alpha \rightarrow 0$ for 
$\beta \rightarrow 0$. This situation 
thus shows 
that the relation between $\alpha$ and $\beta$ is determined by the dynamical mechanism
at work rather being universal as argued by \cite{Hansen_2006}.

From an analytical point of view, by using the hypotheses that 
both the coarse-grained phase-space
density and the density profile being a power law in distance,
allowing for the possibility that 
the velocity distribution is not isotropic
   and the 
empirical linear  relation between $\alpha$ and $\beta$,
it is possible to solve
the Jeans equations analytically and extract the
relevant statistical information of the system  \citep{Dehnen_2005}. 
However, a purely power-law, 
coarse-grained, phase-space density 
approximates well the observed behavior only for the 
case of a bottom-up dynamics but not for the top-down
case. 


 \section{Conclusions}
 \label{sect:concl}

Two competing  processes  work to determine the 
dynamical evolution of finite  initially spherical and cold 
self-gravitating 
systems. On one hand they undergo a global (top-down) collapse  driven by their own rapidly varying gravitational field, and, on the other hand, internal density fluctuations 
 lead to formation of local 
 substructures of growing size through a (bottom-up) 
 aggregation process. Therefore, in general, there is 
 a sort of competition between a top-down  and  a bottom-up  mean-field 
 collisionless dynamics. Anyway, in both cases collisional effects 
 are negligible because of their much longer timescale with respect to that giving rise to  QSS.

 The properties of the QSS formed depend on the evolutionary 
 paths they have followed, and thus on which of the above-mentioned two
 mechanisms prevails during the relaxation from the 
 out-of-equilibrium IC to the quasi-stationary configuration. 
In particular, the dynamics is different depending 
on the type of correlation properties between initial density perturbations.
When the amplitude of initial fluctuations is small, a global collapse takes place and 
the system  relaxes into a QSS in a very short time: 
the signature of this process
is a wide energy exchange between particles. On the other hand, 
in case of large initial fluctuations, the
bottom-up aggregation process becomes predominant over the global collapse 
and, so, clustering at small scales builds up
larger and larger substructures halting the global collapse. That is, the  fragmentation  into large and growing substructures inhibits the occurrence of a large variation of the overall system size and, consequently, the particle energy distribution only moderately changes.

We considered a family of simple IC representing
 isolated, spherical, and cold distributions of particles with different spectra of initial density fluctuations.
By varying the initial amplitude of initial density perturbations, we find that it is possible to 
select the mechanism through which the out-of-equilibrium IC
are driven to form a QSS.

As we said above, for the case of a top-down (monolithic) collapse, which occurs whenever the amplitude of initial density fluctuations are small, 
the particle energy distribution changes significantly in a rapid interval of time centered around the time of maximum system contraction (essentially the free-fall time):
such variation is given by the interplay of the finite size 
of the system with the growth of density perturbations during the collapse.
In this situation, the QSS are characterized by a compact core, which 
contains a significant fraction of the system mass and shows
an almost isotropic velocity distribution.
The core is surrounded by a low-density region in which orbits are 
very elongated, that is, where the velocity anisotropy parameter 
$\beta$ tends to 1.  Based on an assumption of the validity of the Jeans equation, we
were able to show that the inner region of a system emerging by a violent top-down collapse is characterized by an almost flat density profile. However, the outer power-law decay of density is  $\rho(r) \sim r^{-4}$, a behavior that is 
observed in the numerical experiments of initially cold and uniform systems.

On the other hand, when initial perturbations are of large enough amplitude
then  a QSS is reached through a bottom-up, hierarchical, aggregation process in which
small substructures merge to form larger and larger substructures. 
This process is accelerated when initial density correlations 
are long range. That is, at given initial fluctuations amplitude, the smaller
the power-law index, $n \in (-3,0],$ of the density fluctuation power 
spectrum, $P(k) \sim k^n$, the faster the evolution of the bottom-up mechanism of structure formation.
In this latter situation, the variation of the particle energy distribution
is smaller than in the former and, for this reason, the orbits in the outermost 
regions of the system are less radially elongated. The exponent of the density profile is $0 < \alpha \leq 4$ and
the anisotropy parameter is $0\le \beta <1$. When initial density perturbations are large enough the core-halo structure is not formed. 
In this case the profile is better fitted by a NFW behavior.

We also demonstrated that the halos formed in cosmological 
$N$-body simulations in the standard CDM scenario, although they are not isolated but rather embedded in the tidal field of neighboring structures, show properties similar to QSS obtained in the simple isolated and spherical cases considered in this work, in the case of large enough initial fluctuation amplitudes. In CDM-like cosmologies, density fluctuations are long-range 
correlated (i.e., $P(k) \sim k^{-2}$) and, as said above, this situation  
implies the development of a bottom-up aggregation process rather 
than a top-down scenario through the collapse of large over-densities.
We can thus conclude that  isolated, spherical, and dynamically cold systems with different choices of initial density perturbations amplitude represent a useful tool to study the formation of QSS through a mean-field  collisionless dynamics, both when the clustering proceeds in a bottom-up and in top-down way.  
{The fact that systems emerging from  cosmological environment have similar properties to those emerging from 
isolated IC  a  imply that,  when fluctuations are highly nonlinear, the evolution of a cosmological
halo is well approximated by neglecting tidal interactions with neighboring structures.}

To conclude, we now consider a long-standing observational puzzle that 
can be related to these results. This is 
the {core-cusp} problem,  that is, the well-known
difference between the observed inner density profiles of dark matter in
low-mass galaxies and the density profiles obtained in cosmological
$N$-body simulations. Observations seem to indicate an approximately
constant dark matter density in the inner parts of galaxies (the core), while
CDM halos profiles show instead  a $\sim r^{-1}$ power-law
cusp at short distances \citep{NFW_1997}.
This fact, known as the “core/cusp controversy”, 
stands as  one of the
unsolved problems in small-scale cosmology (see for a recent review
\cite{deblok_2010} and references therein).
Our results suggest that a possible solution to this puzzle 
could be found in the violent origin of the galaxies through something more similar to a monolithic collapse than a bottom-up aggregation process.
 \cite{Benhaiem+Joyce+SylosLabini_2017,Benhaiem+SylosLabini+Joyce_2019,SylosLabini_RCD_DLP_2020} 
provide further discussions
of this specific topic. Concerning the 
latter point, we note that the case for galaxy formation through 
a monolithic collapse has also been  very recently advocated  by  \cite{Peebles_2020}. 

\section*{Acknowledgements}

We are grateful to Lehman Garrison for his valuable assistance in 
explaining us the Abacus database and for providing us an ad-hoc
simulation with high resolution halos catalogs and data.  
The Abacus data are available at {\tt https://lgarrison.github.io/AbacusCosmos/}.
We are grateful to Volker Springel for his help in the use of {\tt Gadget-2}.

\bibliographystyle{aa}

\begin{thebibliography}{57}
\expandafter\ifx\csname natexlab\endcsname\relax\def\natexlab#1{#1}\fi

\bibitem[{Aarseth {et~al.}(1988)Aarseth, Lin, \&
  Papaloizou}]{aarseth_etal_1988}
Aarseth, S., Lin, D., \& Papaloizou, J. 1988, Astrophys. J., 324, 288

\bibitem[{Aguilar \& Merritt(1990)}]{aguilar+merritt_1990}
Aguilar, L. \& Merritt, D. 1990, Astrophys. J., 354, 73

\bibitem[{{Arca-Sedda} \& {Capuzzo-Dolcetta}(2014)}]{ASCD2014}
{Arca-Sedda}, M. \& {Capuzzo-Dolcetta}, R. 2014, Astrophys. J., 785, 51

\bibitem[{{Barnes} {et~al.}(2009){Barnes}, {Lanzel}, \&
  {Williams}}]{barnes_etal_2009}
{Barnes}, E.~I., {Lanzel}, P.~A., \& {Williams}, L.~L.~R. 2009, Astrophys. J.,
  704, 372

\bibitem[{Baushev \& Barkov(2018)}]{Baushev_2018}
Baushev, A. \& Barkov, M. 2018, Journal of Cosmology and Astroparticle Physics,
  2018, 034–034

\bibitem[{Benetti {et~al.}(2014)Benetti, Ribeiro-Teixeira, Pakter, \&
  Levin}]{Benetti_2014}
Benetti, F. P.~C., Ribeiro-Teixeira, A.~C., Pakter, R., \& Levin, Y. 2014,
  Phys. Rev. Lett., 113, 100602

\bibitem[{{Benhaiem} {et~al.}(2017){Benhaiem}, {Joyce}, \& {Sylos
  Labini}}]{Benhaiem+Joyce+SylosLabini_2017}
{Benhaiem}, D., {Joyce}, M., \& {Sylos Labini}, F. 2017, Astrophys.J., 851, 19

\bibitem[{{Benhaiem} {et~al.}(2016){Benhaiem}, {Joyce}, {Sylos Labini}, \&
  {Worrakitpoonpon}}]{Benhaiem_etal_2016}
{Benhaiem}, D., {Joyce}, M., {Sylos Labini}, F., \& {Worrakitpoonpon}, T. 2016,
  Astron.Astrophys., 585, A139

\bibitem[{{Benhaiem} \& {Sylos Labini}(2015)}]{Benhaiem+SylosLabini_2015}
{Benhaiem}, D. \& {Sylos Labini}, F. 2015, Mon.Not.R.Astron.Soc., 448, 2634

\bibitem[{{Benhaiem} \& {Sylos Labini}(2017)}]{Benhaiem+SylosLabini_2017}
{Benhaiem}, D. \& {Sylos Labini}, F. 2017, Astron.Astrophys., 598, A95

\bibitem[{{Benhaiem} {et~al.}(2019){Benhaiem}, {Sylos Labini}, \&
  {Joyce}}]{Benhaiem+SylosLabini+Joyce_2019}
{Benhaiem}, D., {Sylos Labini}, F., \& {Joyce}, M. 2019, Phys.Rev.E, 99, 022125

\bibitem[{Binney \& Knebe(2001)}]{Binney+Knebe_2001}
Binney, J. \& Knebe, A. 2001, Mon. Not. Roy. Astron. Soc., 325, 845

\bibitem[{Binney \& Tremaine(2008)}]{Binney_Tremaine_2008}
Binney, J. \& Tremaine, S. 2008, Galactic Dynamics (Princeton University Press)

\bibitem[{{Blumenthal} {et~al.}(1984){Blumenthal}, {Faber}, {Primack}, \&
  {Rees}}]{Blumenthal_etal_1984}
{Blumenthal}, G.~R., {Faber}, S.~M., {Primack}, J.~R., \& {Rees}, M.~J. 1984,
  Nature, 311, 517

\bibitem[{{Blumenthal} {et~al.}(1982){Blumenthal}, {Pagels}, \&
  {Primack}}]{Blumenthal_etal_1982}
{Blumenthal}, G.~R., {Pagels}, H., \& {Primack}, J.~R. 1982, Nature, 299, 37

\bibitem[{Boily {et~al.}(2002)Boily, Athanassoula, \& Kroupa}]{boily_etal_2002}
Boily, C., Athanassoula, E., \& Kroupa, P. 2002, Mon. Not. R. Astr. Soc., 332,
  971

\bibitem[{{Boily} \& {Athanassoula}(2006)}]{boily+athanassoula_2006}
{Boily}, C.~M. \& {Athanassoula}, E. 2006, Mon. Not. R. Astr. Soc., 369, 608

\bibitem[{Bond {et~al.}(1982)Bond, Szalay, \& Turner}]{Bond_etal_1982}
Bond, J.~R., Szalay, A.~S., \& Turner, M.~S. 1982, Phys. Rev. Lett., 48, 1636

\bibitem[{Campa {et~al.}(2014)Campa, Dauxois, Fanelli, \&
  Ruffo}]{Campa_etal_2014}
Campa, A., Dauxois, T., Fanelli, D., \& Ruffo, S. 2014, Physics of Long-Range
  Interacting Systems (Oxford)

\bibitem[{{Capuzzo-Dolcetta}(2019)}]{RCDbook}
{Capuzzo-Dolcetta}, R.~A. 2019, {Classical Newtonian Gravity} (Springer
  International Publishing)

\bibitem[{Dauxois {et~al.}(2002)Dauxois, Ruffo, Arimondo, \&
  Wilkens}]{Dauxois_etal_2002}
Dauxois, T., Ruffo, S., Arimondo, E., \& Wilkens, M. 2002, Dynamics and
  Thermodynamics of Systems with Long-Range Interactions: An Introduction
  (Berlin, Heidelberg: Springer Berlin Heidelberg), 1--19

\bibitem[{{De Blok}(2010)}]{deblok_2010}
{De Blok}, W.~J.~G. 2010, Advances in Astronomy, 2010, 789293

\bibitem[{{Dehnen}(1993)}]{Dehnen_1993}
{Dehnen}, W. 1993, Mon.Not.R.Astron.Soc., 265, 250

\bibitem[{Dehnen \& McLaughlin(2005)}]{Dehnen_2005}
Dehnen, W. \& McLaughlin, D.~E. 2005, Monthly Notices of the Royal Astronomical
  Society, 363, 1057–1068

\bibitem[{Diemand {et~al.}(2004)Diemand, Moore, Stadel, \&
  Kazantzidis}]{diemand_etal_2004}
Diemand, J., Moore, B., Stadel, J., \& Kazantzidis, S. 2004, Mon. Not. Roy.
  Astron. Soc., 348, 977

\bibitem[{Garrison {et~al.}(2018)Garrison, Eisenstein, Ferrer, Tinker, Pinto,
  \& Weinberg}]{Garrison_2018}
Garrison, L.~H., Eisenstein, D.~J., Ferrer, D., {et~al.} 2018, The
  Astrophysical Journal Supplement Series, 236, 43

\bibitem[{Garrison {et~al.}(2019)Garrison, Eisenstein, \&
  Pinto}]{Garrison_2019}
Garrison, L.~H., Eisenstein, D.~J., \& Pinto, P.~A. 2019, Monthly Notices of
  the Royal Astronomical Society, 485, 3370–3377

\bibitem[{Hansen {et~al.}(2006)Hansen, Moore, \& Stadel}]{Hansen_2006c}
Hansen, S., Moore, B., \& Stadel, J. 2006, EAS Publications Series, 20, 33–36

\bibitem[{{Hansen}(2004)}]{Hansen_2004}
{Hansen}, S.~H. 2004, Monthly Notices of the Royal Astronomical Society, 352,
  L41

\bibitem[{Hansen \& Moore(2006)}]{Hansen_2006}
Hansen, S.~H. \& Moore, B. 2006, New Astronomy, 11, 333–338

\bibitem[{Hansen \& Stadel(2006)}]{Hansen_2006b}
Hansen, S.~H. \& Stadel, J. 2006, Journal of Cosmology and Astroparticle
  Physics, 2006, 014–014

\bibitem[{Henon(1964)}]{henon_1964}
Henon, M. 1964, Ann. Astrophys., 27, 1

\bibitem[{{Jeans}(1915)}]{Jea1915}
{Jeans}, J.~H. 1915, Mon.Not.R.Astron.Soc., 76, 70

\bibitem[{Joyce {et~al.}(2009)Joyce, Marcos, \& Sylos~Labini}]{jmsl_2009}
Joyce, M., Marcos, B., \& Sylos~Labini, F. 2009, Mon. Not. R. Astron. Soc.,
  397, 775

\bibitem[{Joyce \& Sylos~Labini(2013)}]{joyce+syloslabini_2013}
Joyce, M. \& Sylos~Labini, F. 2013, Mon. Not. R. Astron. Soc., 429, 1088

\bibitem[{Levin {et~al.}(2008)Levin, Pakter, \& Rizzato}]{levin_etal_2008}
Levin, Y., Pakter, R., \& Rizzato, F. 2008, Phys. Rev., E78, 021130

\bibitem[{Levin {et~al.}(2014)Levin, Pakter, Rizzato, Teles, \&
  Benetti}]{Levin_etal_2014}
Levin, Y., Pakter, R., Rizzato, F.~B., Teles, T.~N., \& Benetti, F.~P. 2014,
  Physics Reports, 535, 1 , nonequilibrium statistical mechanics of systems
  with long-range interactions

\bibitem[{{Lynden-Bell}(1967)}]{lyndenbell_1967}
{Lynden-Bell}, D. 1967, Mon. Not. R. Astr. Soc., 136, 101

\bibitem[{{Merritt} \& {Aguilar}(1985)}]{merritt+aguilar_1985}
{Merritt}, D. \& {Aguilar}, L.~A. 1985, Mon. Not. R. Ast. Soc, 217, 787

\bibitem[{{Navarro} {et~al.}(1997){Navarro}, {Frenk}, \& {White}}]{NFW_1997}
{Navarro}, J.~F., {Frenk}, C.~S., \& {White}, S. D.~M. 1997, Astrophys.J., 490,
  493

\bibitem[{{Navarro} {et~al.}(2004){Navarro}, {Hayashi}, {Power}, {Jenkins},
  {Frenk}, {White}, {Springel}, {Stadel}, \& {Quinn}}]{Navarro_2004}
{Navarro}, J.~F., {Hayashi}, E., {Power}, C., {et~al.} 2004, Monthly Notices of
  the Royal Astronomical Society, 349, 1039

\bibitem[{Padmanabhan(1990)}]{Padmanabhan:1989gm}
Padmanabhan, T. 1990, Phys. Rept., 188, 285

\bibitem[{Peebles(1980)}]{Peebles_1983}
Peebles, P. J.~E. 1980, {The Large-Scale Structure of the Universe} (Princeton
  University Press)

\bibitem[{Peebles(2020)}]{Peebles_2020}
Peebles, P. J.~E. 2020, 
arXiv:2005.07588

\bibitem[{Roy \& Perez(2004)}]{roy+perez_2004}
Roy, F. \& Perez, J. 2004, Mon. Not. R. Astr. Soc., 348, 62

\bibitem[{Sahni \& Coles(1995)}]{sahni_95}
Sahni, V. \& Coles, P. 1995, Physics Reports, 262, 1

\bibitem[{{Spera} \& {Capuzzo-Dolcetta}(2017)}]{SpeRCD2017}
{Spera}, M. \& {Capuzzo-Dolcetta}, R. 2017, Astrophys. Space Sci., 362, 233

\bibitem[{{Springel}(2005)}]{Springel_2005}
{Springel}, V. 2005, Mon.Not.R.Astron.Soc., 364, 1105

\bibitem[{{Sylos Labini}(2012)}]{syloslabini_2012}
{Sylos Labini}, F. 2012, Mon. Not. R. Astron. Soc., 423, 1610

\bibitem[{{Sylos Labini}(2013{\natexlab{a}})}]{syloslabini_2013b}
{Sylos Labini}, F. 2013{\natexlab{a}}, Astron. Astrophys., 552, A36

\bibitem[{{Sylos Labini}(2013{\natexlab{b}})}]{syloslabini_2013}
{Sylos Labini}, F. 2013{\natexlab{b}}, Mon. Not. R. Astron. Soc., 429, 679

\bibitem[{{Sylos Labini} {et~al.}(2015){Sylos Labini}, {Benhaiem}, \&
  {Joyce}}]{SylosLabini+Benhaiem+Joyce_2015}
{Sylos Labini}, F., {Benhaiem}, D., \& {Joyce}, M. 2015, Mon.Not.R.Astron.Soc.,
  449, 4458

\bibitem[{{Sylos Labini} {et~al.}(2020){Sylos Labini}, {Pinto}, \&
  {Capuzzo-Dolcetta}}]{SylosLabini_RCD_DLP_2020}
{Sylos Labini}, F., {Pinto}, L.~D., \& {Capuzzo-Dolcetta}, R. 2020, Physical
  Review E in the press, arXiv:2008.02605

\bibitem[{Taylor \& Navarro(2001)}]{Taylor_2001}
Taylor, J.~E. \& Navarro, J.~F. 2001, The Astrophysical Journal, 563, 483

\bibitem[{Theis \& Spurzem(1999)}]{theis+spurzem_1999}
Theis, C. \& Spurzem, R. 1999, Astron. Astrophys., 341, 361

\bibitem[{van Albada(1982)}]{vanalbada_1982}
van Albada, T. 1982, Mon. Not. R. Astr. Soc., 201, 939

\bibitem[{{Worrakitpoonpon}(2015)}]{worrakitpoonpon_2014}
{Worrakitpoonpon}, T. 2015, Mon. Not. R. Astr. Soc., 466, 1335

\end{thebibliography}

\end{document}